\documentclass[twocolumn,aps,prb,showpacs]{revtex4}
\usepackage{graphicx}

\begin{document}
\title{Crystal truncation rods in kinematical and dynamical x-ray diffraction theories}
\author{Vladimir M. Kaganer}
\affiliation{Paul-Drude-Institut f\"{u}r Festk\"{o}rperelektronik,
Hausvogteiplatz 5--7, D--10117 Berlin, Germany}

\begin{abstract}
Crystal truncation rods calculated in the kinematical approximation
are shown to quantitatively agree with the sum of the diffracted
waves obtained in the two-beam dynamical calculations for different
reflections along the rod. The choice and the number of these
reflections are specified. The agreement extends down to at least
$\sim 10^{-7}$ of the peak intensity. For lower intensities, the
accuracy of dynamical calculations is limited by truncation of the
electron density at a mathematically planar surface, arising from
the Fourier series expansion of the crystal polarizability.
\end{abstract}

\date{\today}



\maketitle

The planar surface of a semi-infinite crystal gives rise to a
diffraction pattern consisting of lines normal to the surface and
passing through the bulk diffraction peaks. The intensity
distribution along these lines, called crystal truncation rods
(CTRs), contains information on the surface
structure.\cite{robinson86,vlieg89} CTR calculations, commonly
performed in the kinematical (single scattering) approximation, are
the basis of x-ray surface crystallography.\cite
{feidenhansl89,robinson91,robinson92,vlieg00} The high-intensity CTR
regions close to the Bragg reflections that require dynamical
(multiple scattering)\ calculations are not sensitive to surface
structure and are usually excluded from considerations in the
surface structure determination studies.

On the other hand, many studies of crystalline films and multilayers
are based on dynamical diffraction theory and are mostly restricted
to the close vicinity of the Bragg peaks.\cite
{BowenTanner98,fewster00,pietsch:book04} The substrate peak position
is commonly used as a reference to study the film relaxation. The
intensity of the substrate peak can also be used as a reference to
obtain the film structure factor on an absolute scale to determine,
e.g., ordering in sublattices.\cite{jenichen05} The thinner the
film, the larger the part of the CTR that needs to be analyzed.
Studies of very thin (a few atomic layers) films require the
analysis of the whole CTR. The problem of an accurate calculation of
the CTR intensity in the whole wave vector range has been the
subject of a number of investigations,\cite
{afanasev86,colella91,caticha93,caticha94,gau95,litzman99,holy03,podorov06,pavlov07}
but a conclusive recipe has not been given yet. It is well
established that the two-beam dynamical diffraction theory is very
accurate in the vicinity of the respective Bragg reflection but
fails in the middle between two reflections. The resolution of the
latter problem was attempted by appropriate multibeam
calculations.\cite{colella91,litzman99,holy03} Recently, Pavlov
\textit{et al.}\cite{pavlov07} proposed to calculate the diffracted
intensity in a wide angular range by summing up the amplitudes of
the diffracted waves for different reflections, each amplitude being
the solution of the corresponding two-beam diffraction problem.

In the present paper, I show that the sum of the solutions of the
two-beam diffraction problems for the reflections along a CTR
coincides with the kinematical formula. The number and the choice of
reflections that have to be included in the summation are specified.
Dynamical and kinematical calculations disagree in the close
vicinity of the Bragg reflections where the dynamical solution is
correct while the kinematical one diverges, and in the regions where
the diffracted intensity is less than $\sim 10^{-7}$ of the peak
intensity. In this latter region, the dynamical calculations are
less accurate. The error in the dynamical calculations originates
from the Fourier expansion of the electron density over the
reciprocal lattice vectors, so that it cannot be remedied by a more
accurate multibeam dynamical theory.

The analysis is based on the equality
\begin{equation}
\frac{iF_{hkL}}{\exp (2\pi iL)-1}=\frac{1}{2\pi }\sum_{l=-\infty }
^{\infty } \frac{F_{hkl}}{L-l}.  \label{eq1}
\end{equation}
It is derived below and the limits of its validity are established.
Here $hkl $ are the integer Miller indices of the reflections ($h$
and $k$ correspond to the directions in the surface plane and $l$ is
along the surface normal), $L$ is the continuous coordinate along a
crystal truncation rod ($L=qa/2\pi $, where $q$ is the $z$-component
of the momentum transfer and $a$ is the lattice spacing in
$z$-direction), and $F_{hkL}=\sum_{j}f_{j}\exp [2\pi
i(hx_{j}+ky_{j}+Lz_{j})/a]$ is the structure factor of the unit cell
(the sum is taken over all atoms in the unit cell, $f_{j}$ is the
atomic scattering factor of the $j$th atom and $x_{j},y_{j},z_{j}$
are its coordinates). It will be shown below that Eq.\ (\ref{eq1})
is valid if and only if the electron density obtained by the back
Fourier transform of $F_{hkl}$ is entirely contained inside the unit
cell. This latter requirement results in the cutting of the atomic
electron densities at the surface.

The left-hand side of Eq.\ (\ref{eq1}) is proportional to the
scattering amplitude of a semi-infinite crystal in the kinematical
approximation.\cite
{robinson86,vlieg89,feidenhansl89,robinson91,robinson92} The
kinematical theory for semi-infinite crystals\cite{pietsch:book04}
provides the remaining prefactors. The amplitude of the scattered
wave is
\begin{equation}
E^{\mathrm{kin}}(hkL)=\frac{\lambda r_{e}}{a^{2}\gamma
_{\mathrm{out}}} \frac{iF_{hkL}}{\exp (2\pi iL)-1}.  \label{eq1a}
\end{equation}
Here $\lambda $ is the wavelength, $r_{e}$ is the classical radius
of the electron, and $\gamma _{\mathrm{out}}=\sin \Phi
_{\mathrm{out}}$, where $\Phi _{\mathrm{out}}$ is the angle between
the diffracted wave and the crystal surface. A similar factor for
the incidence wave, $\gamma _{\mathrm{in}}=\sin \Phi
_{\mathrm{in}}$, where $\Phi _{\mathrm{in}}$ is the angle between
the incident beam and the surface, will be used later. The amplitude
of the incident wave is taken equal to 1.

Each term of the sum on the right side of Eq.\ (\ref{eq1}), after
having been multiplied by the same factor $\lambda
r_{e}/(a^{2}\gamma _{\mathrm{out}})$, is the asymptotic expression
of the two-beam dynamical diffraction theory for angular deviations
from the Bragg peak that are much larger than the dynamical peak
width. Within dynamical theory, the amplitude of the diffracted wave
at large deviations from the Bragg peak for a given reflection
$\mathbf{H}$ is given by $E_{\mathbf{H}} = \chi
_{\mathbf{H}}/[2\gamma _{\mathrm{out}}(\gamma _{\mathrm{in}} +
\gamma _{\mathrm{out}}+\psi _{\mathbf{H}})]$, where $\chi
_{\mathbf{H}}$ is the Fourier component of the polarizability and
$\psi _{\mathbf{H}} = H_{z}/\kappa $ (here $\kappa =2\pi /\lambda $
is the wave vector in vacuum). The position along the CTR
$L=(a/\lambda ) (\gamma _{\mathrm{in}} + \gamma _{\mathrm{out}})$ is
expressed through the incidence and the exit angles, while the
position of the Bragg peak on the CTR is $l=-(a/\lambda )\psi
_{\mathbf{H}}$. Substitution of these expressions gives
$E_{\mathbf{H}} = \lambda r_{e}F_{\mathbf{H}}/[2\pi a^{2}\gamma
_{\mathrm{out}}(L-l)]$, which is equal to the corresponding term in
Eq.\ (\ref{eq1}) multiplied by the factor given above. Then, the sum
of the amplitudes of diffracted waves calculated in the two-beam
dynamical theory
\begin{equation}
E^{\mathrm{dyn}}(hkL)=\sum\nolimits_{l}E_{hkl}^{\mathrm{dyn}}  \label{eq2}
\end{equation}
for all reflections $hkl$ along a CTR\ $hk$ coincides with the
kinematical solution. For non-grazing incidence and exit, the
resulting amplitude depends on the $z$-component of momentum
transfer $q=2\pi L/a$, rather than the incident and the exit angles
$\Phi _{\mathrm{in}}$ and $\Phi _{\mathrm{out}}$ separately. Pavlov
{\em et al.}\cite{pavlov07} have arrived to an equivalent formula by
solving the Takagi-Taupin equations.

In the close vicinity of each reflection $hkl$, for angular
deviations comparable to the Darwin width, the amplitudes of the
other reflections are much smaller (on the order of the
polarizability $\chi \sim 10^{-5}$) than that of the actual
reflection, and Eq.\ (\ref{eq2}) can be applied, keeping in mind
that all other terms can be neglected compared to
$E_{hkl}^{\mathrm{dyn}}$. At larger angular deviations, Eq.\
(\ref{eq2}) is the first order (in $\chi $) approximate solution of
the multibeam dynamical diffraction problem that includes all Bragg
reflections along a given CTR. Such an approximation is valid since
at most one Bragg reflection occurs at once at the diffraction
condition when moving along the CTR. Hence, Eq.\ (\ref{eq2})
describes the intensity distribution along a CTR around Bragg peaks
and far from them, down to intensities $\sim \chi $ of the peak
intensity, which covers almost the whole CTR. Eq.\ (\ref{eq2})\
fails at the order of $\chi ^{2}$. Before ascertaining the problems
arising at such low intensities, let us consider the CTR
calculations based on Eq.\ (\ref{eq2}).

I have implemented the dynamical diffraction equations in the
formulation of Stepanov \textit{et al.}\cite{stepanov98} Since Eq.\
(\ref{eq2})\ involves the summation of the scattering amplitudes, a
proper treatment of the phases of the Fourier components $\chi
_{hkl}$ of the polarizability is essential. The structure factor
calculations require the origin of the unit cell to be chosen once
for all reflections. This is not so in the programs by Sergey
Stepanov\cite{stepanov:www} that choose the origin separately for
each reflection to minimize the phase of $\chi _{hkl}$. Such a
choice is correct as long as a single reflection is involved in the
calculations. When working with the amplitudes of several
reflections, Eq.\ (\ref{eq2}), this is not appropriate. Also, for
layered structures, the interference pattern depends on the relative
displacements of the crystal lattices of different layers, which
requires a common origin for the calculation of all $\chi
_{hkl}$'s.\cite{jenichen05} The requirement of a common origin is
usual in surface x-ray structure analysis but not appreciated in the
dynamical diffraction calculations.

Equally important, but probably less evident, is the choice of the
atomic positions in the unit cell with respect to the surface. The
surface is commonly taken at $z=0$, and one of the atoms is placed
at the origin. If so, the surface cuts half of this atom (and other
atoms at the same level), which leads to wrong CTR\ intensities. For
example, with the surface at $z=0$, the Ga and As atoms in the bulk
GaAs unit cell cannot be placed at the depths 0, 1/4, 1/2, 3/4.
Rather, they have to be put at the depths 1/8, 3/8, 5/8, 7/8. Only
this choice leads to an agreement between dynamical and kinematical
calculations. Note that this choice of the atomic positions is
irrelevant to the kinematical formula (\ref{eq1a}), since the
kinematical calculation, based on the summation over atoms and unit
cells, contains only one structure factor $F_{hkL}$ (with the
continuous parameter $L$), the phase of which is lost when
calculating the intensities. In contrast, the relative phases of the
structure factors $F_{hkl}$ (with integer $l$'s) depend on the
choice of atomic positions.

The choice of the reflections that have to be included in the sum
(\ref{eq2}) follows from Eq.\ (\ref{eq1}). Since $L$ starts from
zero, positive and negative $l$'s are equally important. Hence, the
sum should contain the solutions of the two-beam dynamical problems
in both reflection (the Bragg case) and transmission (the Laue
case). Although the dynamical diffraction equations are formulated
in Ref.\ \onlinecite{stepanov98} for the reflection case, they do
not require any change to include the transmission case, since the
diffracted wave is anyway directed back into vacuum. In other words,
the reflections that formally correspond to the Laue case (the
diffracted wave propagates into the crystal when the Bragg condition
is satisfied) become the Bragg case reflections because the
deviation from the Bragg condition is so large that the diffracted
wave propagates into the vacuum. The number of equations to be
included in the sum (\ref{eq1}) or (\ref{eq2}) depends on the decay
of the structure factors $F_{hkl}$ with increasing $l$. The atomic
scattering factors $f(s)$ are decreasing functions of $s=\sin \theta
/\lambda $, where $\theta $ is the Bragg angle of the respective
reflection, with $s\approx 2$~\AA $^{-1}$ as a characteristic scale
for this decay. Then, for reflections with $l\gg h,k$ the lattice
spacing is $d=a/l$ and, from Bragg's law $s=1/2d$, reflections up to
$l_{\max }\approx 4a$ have to be included (here $a$ is in
angstroms). Hence, a reasonable estimate is $l_{\max }\approx 25$.
In fact, the calculations below show that $l_{\max }\approx 10$ is
sufficient for practical calculations.

\begin{figure*}[t!]
\noindent \includegraphics[width=\textwidth]{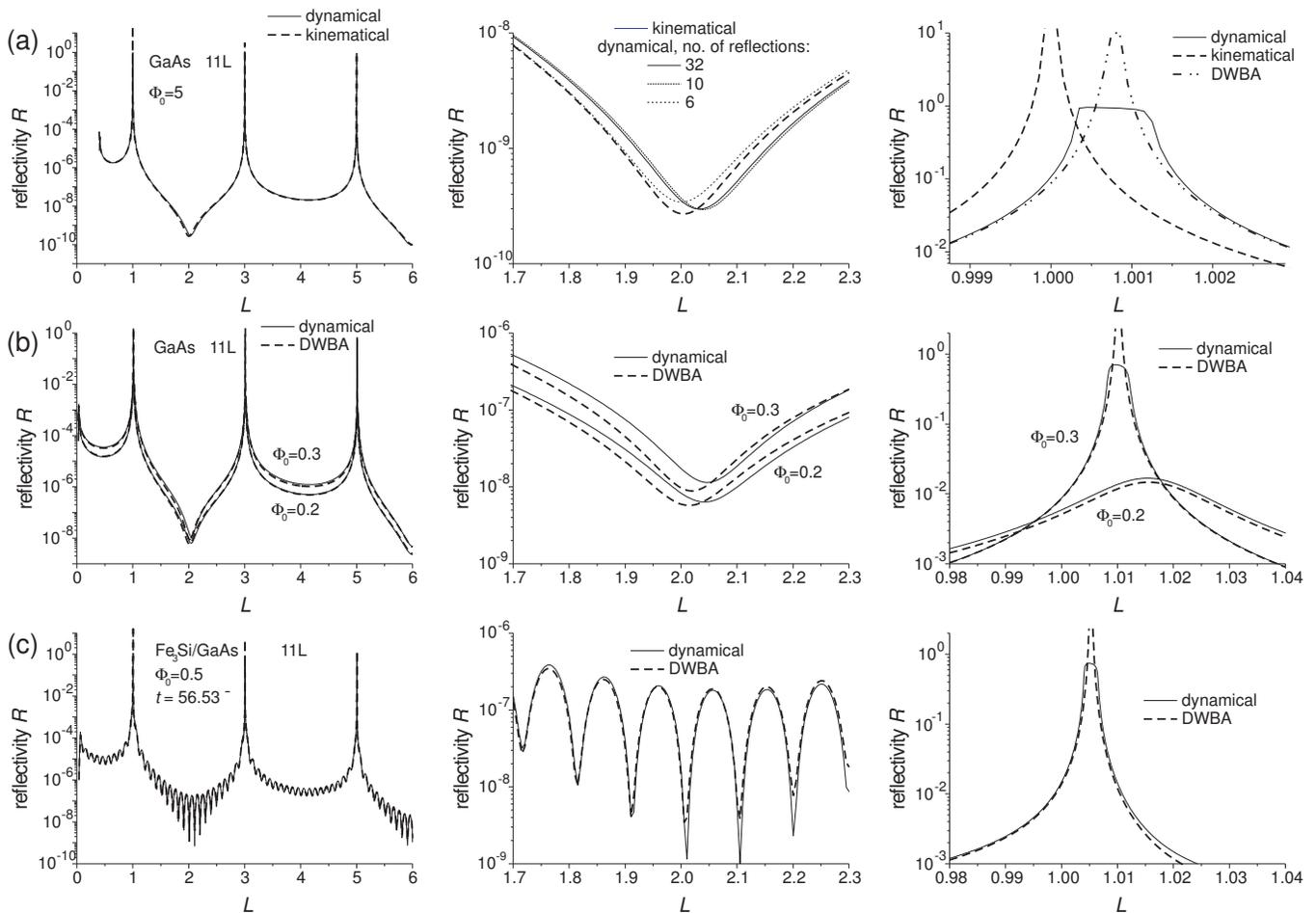} \caption{
Crystal truncation rods $11L$ from a bulk-terminated GaAs(001)
crystal at (a) non-grazing incidence angle $\Phi_0=5^\circ$ and (b)
grazing incidence angles $\Phi_0=0.3^\circ$ and $\Phi_0=0.2^\circ$,
above and below the critical angle $\alpha_c=0.24^\circ$. (c) CTR
$11L$ from a 56.53~\AA~thick Fe$_3$Si film on GaAs(001) at an
incidence angle $\Phi_0=0.5^\circ $. The two right panels magnify
the curves of the left panel at the intensity minimum
(bulk-forbidden reflection 112) and close to the maximum (bulk
reflection 111). Dynamical, kinematical, and distorted-wave Born
approximations are compared.} \label{CTR}
\end{figure*}

\begin{figure}[tbh]
\noindent \includegraphics[width=6cm]{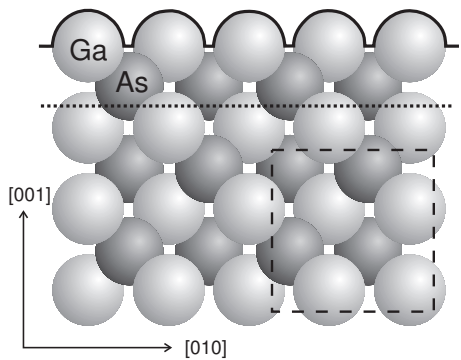} \caption{View of a
truncated GaAs bulk structure in the [100] direction. Atoms are
represented by balls with their radii equal to the covalent radius
of the respective element. A unit cell marked by broken lines is
chosen to minimize the overlap of the electron densities from atoms
of different unit cells. The dotted straight line represents a cut
by a mathematically planar (001) surface. The solid line is a
physical (001) surface terminated by atoms. } \label{GaAs}
\end{figure}

Figure\ \ref{CTR}(a) compares kinematical and dynamical calculations
of the CTR\ $11L$ from  a hypothetical bulk-terminated GaAs(001)
crystal. The reflectivities $R=(\gamma _{\mathrm{out}}/\gamma
_{\mathrm{in}})\left| E(hkL)\right| ^{2}$ are plotted. A non-grazing
incidence angle $\Phi _{0}=5^{\circ }$ is chosen to ensure
kinematical diffraction conditions. In the left panel, presenting
the whole CTR, the two calculated curves are almost
indistinguishable. At the 111 bulk reflection (the right panel) the
kinematical curve is slightly shifted with respect to the dynamical
one because of refraction at the surface, and diverges at the Bragg
position. A small but important discrepancy between the curves is
revealed in the middle panel that enlarges the region of the
intensity minimum, close to the bulk-forbidden reflection 112.
Dynamical calculations are performed with different numbers of terms
in the sum (\ref {eq2}). One can see from the plot that 10 terms
(i.e., the sum of two-beam dynamical calculations for the
reflections $11l$ with odd $l$ in the range $-9\leq l\leq 9)$ are
sufficient for the convergence of the series. However, the sum
(\ref{eq2}) does not converge to the kinematical solution even if
the number of the involved reflections is increased further. The
origin of the disagreement is in the Fourier expansion of the
electron density over the reciprocal lattice vectors $hkl$. This
expansion is certainly correct for an infinite bulk crystal that
consists of a periodic repetition of the unit cells. However, if a
unit cell is cut out of the crystal, its electron density does not
coincide with the sum of electron densities of the atoms that belong
to this unit cell. Figure \ref{GaAs} illustrates this statement.
Here the atoms of the GaAs bulk structure are represented by balls
with radii equal to their covalent radii. The broken line selects
one unit cell. Its position is chosen to minimize the overlap of the
electron densities belonging to the atoms of different unit cells.
Still, parts of the electron density of atoms of the chosen unit
cell are out of that unit cell, and parts of atoms from neighboring
unit cells fall into the chosen unit cell. These do not affect the
bulk structure factor calculations, since the unit cells are
periodically repeated. However, the dynamical Bragg diffraction from
a semi-infinite crystal implies a truncation of the sum over the
unit cells, which is equivalent to a rigid truncation of the
electron density by a plane (dotted line in Fig.\ \ref{GaAs}). Such
a cut of the electron density of an infinite crystal by a plane
removes parts of the top atoms and artificially adds the parts of
the atoms belonging to the next layer. In contrast, the physical
surface is rough on that scale since it contains the entire
densities of the atoms of the top layer, as shown by the thick line
in Fig.\ \ref{GaAs}.

\begin{figure*}[tbh]
\noindent \includegraphics[width=\textwidth]{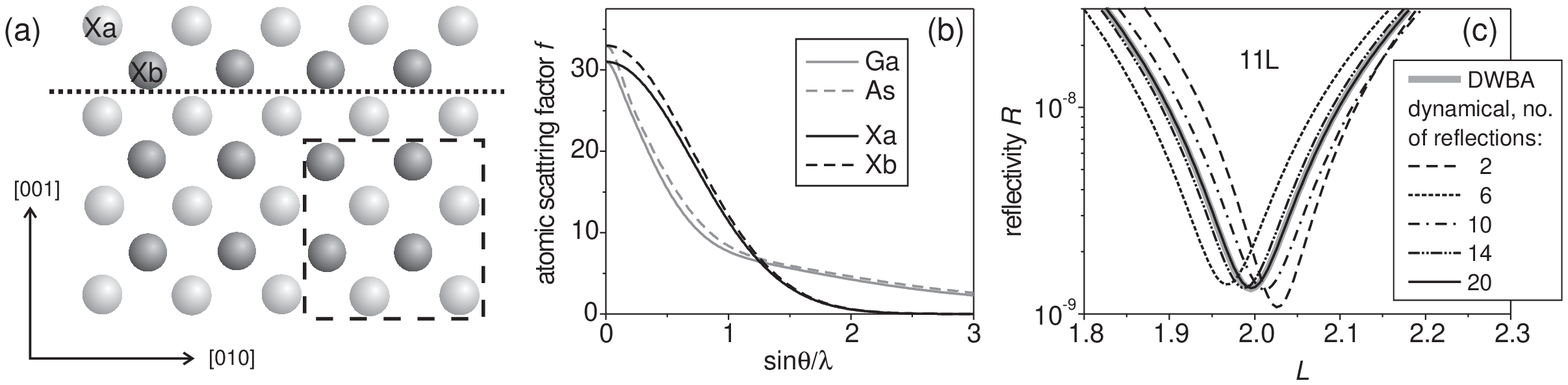} \caption{ (a)
An artificial crystal XaXb with the same structure and the lattice
parameter as GaAs but with atomic sizes that are two times smaller.
The unit cell (broken lines) and a planar surface (dotted line) can
be cut between atoms. (b) Atomic scattering factors attributed to
atoms Xa and Xb. The functions $f(\sin \theta/\lambda)$ are chosen
to be Gaussians with widths approximately two times larger than
those of Ga and As. (c) The CTR $11L$ from the XaXb semi-infinite
crystal in the vicinity of $L \approx 2$ calculated within
distorted-wave Born approximation (DWBA) and dynamically, by Eq.\
(\ref{eq2}), with different numbers of involved reflections. As the
number of reflections is increased, the dynamical curves converge to
the DWBA results. The incidence angle is $\Phi_0=0.5^\circ$.}
\label{XaXb}
\end{figure*}

The difference in these electron densities does not play a role in
the vicinity of Bragg peaks, where many atoms contribute to
diffraction, but becomes essential between Bragg peaks, where the
surface atoms provide the main contribution to the diffracted wave.
To provide an additional proof that this effect is the only source
of discrepancy between the dynamical and kinematical calculations, I
have repeated the same calculations for artificial atoms two times
smaller in size placed in the same lattice, Fig.\ \ref{XaXb}. Such a
crystal is not, in fact, physical, since the long-range attractive
forces between atoms are not compensated by the short-range
repulsion when the distances between atoms exceed the sum of their
covalent radii by a factor of 2. The atomic scattering factors for
the artificial atoms Xa and Xb are chosen to be Gaussians with
widths approximately two times larger than the $f(s)$ functions for
Ga and As atoms, Fig.\ \ref{XaXb}(b). The convergence of sum
(\ref{eq1}) or (\ref{eq2}) is fairly slow, Fig.\ \ref{XaXb}(c). The
sum of dynamical amplitudes converge to the distorted-wave Born
approximation, described below, when 20 terms are included. Each
term is a solution of the two-beam dynamical diffraction problem for
a reflection $11l$ with odd $l$ from $-17$ to $+21$.

This result discourages other improvements of the dynamical theory,
in particular a complete solution of the multibeam diffraction
problem, that could improve the accuracy compared to Eq.\
(\ref{eq2}). As long as a Fourier expansion over reciprocal lattice
vectors is employed, the cut of the electron density shown in Fig.\
\ref{GaAs} persists and does not provide an accuracy that is better
than given by Eq.\ (\ref{eq2}), i.e., an accuracy of the order of
$\sim 10^{-7}$ with respect to the peak intensity. Fortunately,
lower intensities quite rarely arise in experimental CTR studies and
are additionally masked by surface reconstruction and roughness.

Equation (\ref{eq2}) can be directly applied to grazing incidence
diffraction with the same reasoning as above:\ in the vicinity of a
Bragg peak, corrections due to other reflections are negligible, and
in the remaining part of a CTR the sum (\ref{eq2}) is the first
order (over $\chi $) perturbation solution of the multibeam
diffraction problem. The kinematical approximation is extended, for
the grazing incidence diffraction case, as the distorted-wave Born
approximation (DWBA).\cite
{pietsch:book04,sinha88,dmitrien87,kag+step95} The DWBA\ formulation
in terms of the reciprocity theorem in electrodynamics is most
straightforward.\cite {dmitrien87,kag+step95} In the zeroth order,
the scattering problem is solved for a uniform medium having the
same polarizability $\chi _{0}$ as the crystal under investigation.
The scattering problem is solved twice, with the waves incident on
the surface under the incidence angles $\Phi _{\mathrm{in}}$ and
$\Phi _{\mathrm{out}}$, respectively. The solution of each problem
in the medium consists of two plane waves with the amplitudes
$D_{i}$ ($i=1,2$), corresponding to the transmitted and the
specularly reflected waves in vacuum. A convenient way to find these
solutions is to reduce the 4$\times 4$ matrices of the dynamical
diffraction problem\cite{stepanov98} to 2$\times 2$ matrices. In
this way, layered structures with different $\chi _{0}$'s can easily
be treated. Then, the kinematical solution is replaced by
\begin{equation}
E^{\mathrm{DWBA}}(\Phi _{\mathrm{in}},\Phi _{\mathrm{out}}) =
\sum_{i,j=1,2}D_{i} ^ {\mathrm{in}}D_{j} ^ {\mathrm{out}}
E^{\mathrm{kin}} (hkL_{ij}). \label{eq3}
\end{equation}
The superscripts ``in'' and ``out'' distinguish the respective
zeroth order solutions. The parameters $L_{ij}=(a/\lambda) (u_{i} ^
{\mathrm{in}} + u_{j}^{\mathrm{out}})$ are obtained from the
parameters $u_{i}=k_{zi}/\kappa $ describing the wave vectors in the
medium (here $\mathbf{k}_{i}$ are the wave vectors of the waves
inside the medium). Complex parameters $L_{ij}$ substitute the real
parameter $L$ in Eq.\ (\ref{eq1a}).

Figure\ \ref{CTR}(b) compares dynamical and DWBA calculations of the
same CTR\ $11L$ for the incidence angles below and above the
critical angle $\alpha _{c}=0.24^{\circ }$. A good overall agreement
is evident from the left panel. The discrepancy at the intensity
minimum (middle panel) has the same nature as above. The right
panels of Fig.\ \ref{CTR}(a,b) show that the DWBA and the dynamical
calculation agree everywhere except within the Darwin width of the
Bragg peak. If the incidence angle is smaller than the critical
angle, the results of the two calculations coincide in that region
as well.

The extension to layered structures is straightforward. The
diffracted beam amplitudes obtained in the two-beam dynamical
calculations\cite{stepanov98} are summed up according to Eq.\
(\ref{eq2}). The kinematical amplitude (\ref {eq1a}) is supplemented
by an additional term arising from the finite sum over the layer's
unit cells, $F_{hkL}^{\prime }[\exp (2\pi iNL)-1]/[\exp (2\pi
iL)-1]$, where $F_{hkL}^{\prime }$ is the structure factor of the
layer unit cell, and $Na$ is the layer thickness. Figure
\ref{CTR}(c) compares dynamical and DWBA calculations for a
Fe$_{3}$Si film on GaAs(001). Fe$_{3}$Si possesses a cubic unit cell
with the same lattice spacing as GaAs, so that there is no mismatch
between the two lattices.\cite{jenichen05} The incidence angle is
taken $\Phi _{0}=0.5^{\circ }$, so that the DWBA corrections to the
kinematical formulas are essential. The film is $N=10$ unit cells
thick ($Na=56.53$ \AA ). A perfect agreement between the two curves
is evident. Even at the intensity minimum (middle panel), the
discrepancy between the two calculations is absent. At the bulk
reflection 111 (right panel), the DWBA solution diverges and cannot
be used within the Darwin width, while the dynamical solution is
correct. The dynamical calculations in Figs. \ref{CTR}(b,c) were
performed by the summation of solutions of 32 two-beam diffraction
problems ($11l$ reflections with odd $l$ in the range $-31\leq l\leq
31$ were calculated).

\begin{figure*}[tbh]
\noindent \includegraphics[width=\textwidth]{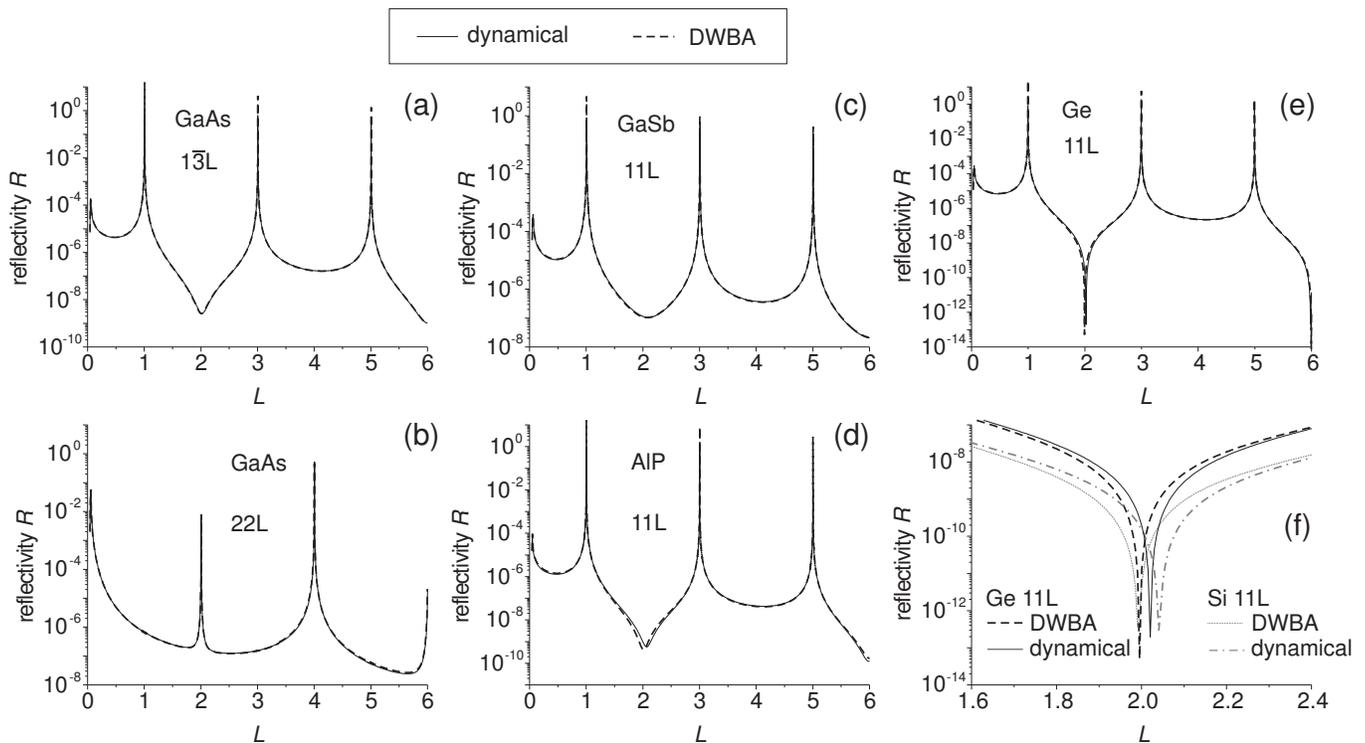}
\caption{Crystal truncation rods from GaAs (a,b), other A$_{\rm
III}$B$_{\rm V}$ compounds (c,d), and elemental semiconductor (e,f)
bulk-truncated (001) surfaces. Dynamical (full lines) and DWBA
(broken lines) calculations are compared. The incidence angle is
$\Phi_0=0.5^\circ$.} \label{otherCTRs}
\end{figure*}

The CTR\ $11L$ analyzed in Fig.\ \ref{CTR} shows a larger
discrepancy between dynamical and kinematical calculations than
other rods from GaAs(001) surface. Figures \ref {otherCTRs}(a,b)
present two other rods. The difference between the two calculations
diminishes for the $1\bar{3}L$ rod and almost disappears for the
$22L$ rod. Further calculations show that the discrepancy decreases
for A$_{\mathrm{III}}$B$_{\mathrm{V}}$ crystals consisting of
elements of two different periods in the periodic table [e.g., GaSb,
see Fig.\ \ref{otherCTRs}(c)] and increases for lighter elements
from the same period [e.g., AlP, see Fig.\ \ref{otherCTRs}(d)]. The
discrepancy remains quite small on the scale of the whole CTR and
hardly can have practical implications. Elemental semiconductors
show more sharp minima at the bulk-forbidden reflection 112, see
Figs.\ \ref{otherCTRs}(e,f). Similarly to the case of compound
semiconductors, silicon as a lighter element shows a larger, as
compared to germanium, discrepancy between dynamical and DWBA
calculations, see Fig.\ \ref{otherCTRs}(f).

Equation (\ref{eq1}) remains to be derived. Consider an arbitrary
complex one-dimensional function $\rho (z)$ that is equal to zero
outside a finite interval $[0,a]$.\ The Fourier integral of the
function $\rho (z)$ therefore involves integration over a finite
interval:
\begin{equation}
F(q)=\frac{1}{a}\int_{0}^{a}\rho (z)\exp (iqz)dz.  \label{eq4}
\end{equation}
On the other hand, the function $\rho (z)$ can be periodically
repeated along the $z$-axis,
\begin{equation}
\rho _{\infty }(z)=\sum_{l=-\infty }^{\infty }\rho (z+la),  \label{eq5}
\end{equation}
and the periodic function $\rho _{\infty }(z)$ can be expanded into
the Fourier series:
\begin{equation}
\rho _{\infty }(z)=\sum_{l=-\infty }^{\infty }F(q_{l})\exp (-iq_{l}a),
\label{eq6}
\end{equation}
where $q_{l}=2\pi l/a$. Let $\Omega (z)$ be a function equal to one
on the interval $[0,a]$ and equal to zero outside this interval.
Then, identically
\begin{equation}
\rho (z)=\rho _{\infty }(z)\Omega (z).  \label{eq7}
\end{equation}
The Fourier integral of the left-hand side of this equation is equal
to $F(q) $. The Fourier integral of the right-hand side is
calculated by performing the integration for each term of the sum:
\begin{eqnarray}
F(q) &=&\sum_{l=-\infty }^{\infty }F(q_{l})\int_{0}^{a}\exp [i(q-q_{l})a]dz
\nonumber \\
&=&[\exp (iqa)-1]\sum_{l=-\infty }^{\infty }\frac{F(q_{l})}{i(q-q_{l})a}.
\label{eq8}
\end{eqnarray}
It is taken into account that $\exp (iq_{l}a)=1$ since $q_l=2\pi
l/a$ with an integer $l$, so that $q_l a$ is a multiple of $2\pi $.
The function $\rho (z)$ can now be identified with the Fourier
component $\rho _{hk}(z)$ in the expansion of the electron density
of a one unit cell thick crystalline slab over the wave vectors of
its two-dimensional reciprocal lattice. Its Fourier transform $F(q)$
is equal to $F_{hkL}$ with $L=qa/2\pi $, and $F(q_{l})$ are equal to
$F_{hkl}$. Equation (\ref{eq8}) reduces to Eq.\ (\ref{eq1}). The
crucial point in the derivation is contained in Eq.\ (\ref{eq7}),
which explicitly requires that the function $\rho (z)$ is restricted
to the interval $[0,a]$.

Hence, the structure factors in Eq.\ (\ref{eq1}) correspond to the
electron density constrained to a unit cell cut out of the crystal,
rather than to the electron density of the atoms whose centers are
inside this unit cell. Parts of the electron density distributions
of the atoms belonging to the chosen unit cell are cut away, while
parts of the atoms from the surrounding unit cells occur in the
chosen unit cell, Fig.\ \ref{GaAs}. The problem can be avoided in
rare cases of layered crystals, e.g. graphite, but is enhanced for
higher-index surfaces. This effect limits the applicability of the
Fourier series expansion of the electron density for the solution of
the diffraction problems. The solution of a multibeam diffraction
problem instead of Eq.\ (\ref{eq2}), as well as corrections to the
dynamical equations that are omitted in Ref.\
\onlinecite{stepanov98}, provide improvements of the order of $\chi
^{2}$, smaller than the effect of rigid truncation. That is why
these corrections are not included in the calculations here.
However, only the regions of extremely low intensity are affected.
Intensities in these regions are sensitive to surface reconstruction
and roughness. Hence, further improvements of the dynamical theory
seem of limited practical impact nowadays.

In conclusion, it is shown that the kinematical calculation (or, at
small incidence angles, the distorted-wave Born approximation)
quantitatively agrees with the sum of the diffracted beam amplitudes
obtained in the two-beam dynamical calculation. The number of Bragg
reflections that have to be included in the dynamical calculations
is estimated from the angular dependence of the atomic scattering
factors and amounts to some tens of reflections. Both transmission
(Laue) and reflection (Bragg) cases have to be included. The
reference unit cell should be chosen to minimize (since it cannot be
completely excluded)\ the cutting of the electron densities of the
top atoms by the surface. The agreement between kinematical and
dynamical calculations can be lost in two regions. In the Darwin
width regions near Bragg reflections the dynamical theory provides
correct intensities while the kinematical theory diverges. In the
regions of very low intensity (below $\sim 10^{-7}$ of the peak
intensity), dynamical theory may fail because of the electron
density truncation by a mathematically flat plane instead of the
physical surface while the kinematical theory remains applicable.

This work has been inspired by discussions at the 8th Conference on
High Resolution X-Ray Diffraction and Imaging (XTOP 2006).


\end{document}